# High temperature magnetic ordering in La$_2$RuO$_5$

S. K. Malik*, Darshan C. Kundaliya‡ and R. D. Kale

Tata Institute of Fundamental Research, Colaba, Mumbai, 400 005, India

Magnetic susceptibility, heat capacity and electrical resistivity measurements have been carried out on a new ruthenate, La$_2$RuO$_5$ (monoclinic, space group P2$_1$/c) which reveal that this compound is a magnetic semiconductor with a high magnetic ordering temperature of 170K. The entropy associated with the magnetic transition is 8.3 J/mole-K – close to that expected for the low spin (S=1) state of Ru$^{4+}$ ions. The low temperatures specific heat coefficient $\gamma$ is found to be nearly zero consistent with the semiconducting nature of the compound. The magnetic ordering temperature of La$_2$RuO$_5$ is comparable to the highest known Curie temperature of another ruthenate, namely, metallic SrRuO$_3$, and in both these compounds the nominal charge state of Ru is 4+.

*Electronic address: skm@tifr.res.in

In the last few years, there has been a great deal of interest in the study of ruthenates which form in a variety of structures ranging from perovskites to pyrochlores. Many of these compounds exhibit remarkable properties arising due to the hybridization of the Ru-4d orbitals and the O-2p orbitals. These properties range from high temperature ferromagnetic ordering [1] in metallic $SrRuO_3$ (Curie temperature, $T_C$=165K), unexpected low temperature superconductivity [2] in metallic $Sr_2RuO_4$ below 1.5 K, non-Fermi liquid behaviour [3] in $La_4Ru_6O_{19}$, etc. We report here the observation of magnetic ordering with a very high ordering temperature of ~170K in a new non-metallic system, $La_2RuO_5$. The magnetism due to 4d electrons is rather rare and the observation of such high ordering temperatures is even more rare.

The $La_2RuO_5$ compound was made by the standard ceramic technique. Stoichiometric amounts of $La_2O_3$ and $RuO_2$ were thoroughly mixed, palletized and sintered at 1150ºC for 48 hours with several intermediate grindings to ensure homogeneous compound formation. Powder x-ray diffraction patterns were obtained using Cu-K$_\alpha$ radiation (Siemens X-ray diffractometer). Magnetic measurements were carried out in the temperature range of 1.8-300K and in applied magnetic fields up to 5.5 Tesla using a Squid magnetometer (MPMS, Quantum Design). The heat capacity measurements in the temperature range of 1.9-300K were made using the relaxation method (PPMS, Quantum Design). Four-probe dc resistivity measurements were also carried out in the temperature range of 300-120K (PPMS, Quantum Design).

The crystal structure of the new ruthenate, $La_2RuO_5$, has been recently solved [4]. This compound is found to crystallize in the monoclinic structure (space group $P2_1/c$, No. 14, Z=4). This structure is not only different than the orthorhombic structure of the other known $R_2RuO_5$ (R=Pr-Gd) compounds [5], but also appears to be unique in the family of oxides. According to ref. 4, the structure can be described as a stacking of $[LaRuO_4]_\infty$ of two-layer thickness and a slab of $[LaO]_\infty$ of 3.4Å thickness.

The Powder x-ray diffraction pattern of $La_2RuO_5$ sample, synthesized by us, could be refined by the Rietveld method in the above-mentioned space group with the starting atomic positions given in ref. 4. The observed and fitted x-ray diffraction patterns are shown in Fig. 1 from which it is very clear that the sample is single phase with in the limits of x-ray detection. The refined lattice parameters are a=9.1684Å, b=5.8263Å, c=7.9437Å, β=100.756° (V=416.88Å$^3$) which are in very good agreement with those reported for the same [4]. The structure of $La_2RuO_5$ has two La sites, five O sites but a unique Ru site. The Ru atoms appear to be located in the center of an octahedral oxygen arrangement. The Ru-O distances range from 1.93Å to 2.13Å. The O-Ru-O angles range from 82° to 95°, which suggests a distorted octahedral coordination for the Ru ions. The shortest Ru-Ru distance is ~3.97Å. As mentioned earlier, the structure of $La_2RuO_5$ is very different from that of other $R_2RuO_5$ (R=Pr-Gd) compounds and so are its magnetic properties.

The magnetization of $La_2RuO_5$ was first measured in a low applied field of 50Oe both in the zero-field-cooled (ZFC) and the field-cooled (FC) states. Figure 2 shows a plot of magnetization (M) versus temperature (T). As temperature is lowered from 300K, a peak followed by a sharp drop in the magnetization is seen at about 170K indicative of a transition to a magnetic state, most likely to an antiferromagnetic state (however, see below). At still lower temperatures, a branching of $M_{FC}$ and $M_{ZFC}$ occurs at about 50K. While $M_{ZFC}$ shows a large increase below this temperature, $M_{FC}$ goes through a peak.

This suggests the possibility of either another magnetic transition or a re-arrangement of magnetic structure. However, it may be mentioned that the presence of a small magnetic impurity (below the limits of x-ray detection), which may lead to the branching of ZFC and FC magnetization curves, cannot be ruled out. The magnetization (M) versus field (H) isotherms, obtained at various temperatures, are shown in Fig. 3. The M-H isotherm at 100K shows a linear behaviour as would be expected for an antiferromagnet. However, the M-H isotherm at 5K shows a small deviation from linearity consistent with the formation/presence of a small ferromagnetic component.

The plot of magnetic susceptibility ($\chi$), measured in a field of 5kOe, versus temperature for $La_2RuO_5$ is shown in Fig. 3 along with a plot of $\chi^{-1}$ versus temperature. Above the magnetic ordering temperature, the susceptibility follows the Curie Weiss behaviour, $\chi=C/(T-\theta_P)$, with an effective magnetic moment, $\mu_{eff}=2.72\mu_B$ and paramagnetic Curie temperature, $\theta_P=-245K$. The large negative value of $\theta_P$ is consistent with an antiferromagnetic ordering in this compound. The nominal charge-state of Ru in $La_2RuO_5$ is 4+ which corresponds to the configuration $4d^4$. The $Ru^{4+}$ ion can have a total spin S=1 or S=2. The crystalline electric field at the Ru site in this compound is of octahedral symmetry. In the presence of crystal-field-splitting, the 4 electrons fill the lowermost 3-fold degenerate $t_{2g}$ orbitals $d_{xy}$, $d_{yz}$ and $d_{xz}$ and hence two electrons must pair to give rise to the low spin state with S=1. The effective magnetic moment value of $2.72\mu_B$ per Ru ion observed in the present study on $La_2RuO_5$ is indeed close to that of the spin only value of S=1 state ($2.83\mu_B$) of $Ru^{4+}$ ion.

It is interesting to note that the magnetic ordering temperature of $La_2RuO_5$ is comparable to the ferromagnetic ordering temperature, $T_C$, of metallic $SrRuO_3$ in both of which the nominal charge state of Ru is 4+. Due to the relatively larger extent of 4d orbitals (compared to that of the 3d orbitals), the occurrence of magnetism from 4d electrons is rather rare. It is even more rare to have a very large magnetic ordering temperature. The magnetic ordering in other $R_2RuO_5$ [R=Pr-Gd] occurs in the 20K range [5] and is attributed to the ordering of the rare earth (R) moments through R-R interaction enhanced by the presence of the 4d electrons. This suggests that the underlying crystal structure may strongly influence the Ru-Ru magnetic interaction. This interaction in $La_2RuO_5$ is expected to be through the super-exchange process involving O-2p orbitals. It has been shown [6] that when neighboring Ru ions are bonded to the same O, then the resulting Ru-Ru interaction is ferromagnetic. However, if Ru ions are not connected via common O ions, the resulting Ru-Ru interaction is antiferromagnetic.

The heat capacity (C) of $La_2RuO_5$ is shown in Fig. 4 as a function of temperature (T). A pronounced peak is seen in the C versus T curve at a temperature of ~161K, which is close to the magnetic ordering temperature obtained through magnetization measurements. However, no anomaly in the heat capacity is seen at 50K where a branching in the $M_{ZFC}$ and $M_{FC}$ curves is observed. In the absence of a suitable reference for estimating the lattice contribution to the specific heat, as a first approximation, the entropy associated with the transition is obtained by first subtracting the smoothened background, consisting of electronic and lattice specific heat contribution, from the total heat capacity and integrating the resulting $C_{mag}/T$ versus T curve (Fig. 5). The entropy so obtained is 8.3 J/mol-K per Ru which is close to the value of Rln(2S+1) of 9.13 J/mol-K for S=1 for the low spin ground state of $Ru^{4+}$ (considering that our procedure slightly overestimates the nonmagnetic contribution and hence slightly underestimates $C_{mag}$ and the associated entropy). Thus magnetic and heat capacity measurements are consistent with the low spin ground state of $Ru^{4+}$ with S=1. However, this result should be taken with some caution since it has been reported that there is a structural transformation concomitant with the magnetic transition (see below) and the contribution to entropy associated with the structural transformation is not known.

To obtain the value of electronic specific heat coefficient, $\gamma$, and the Debye temperature, $\Theta_D$, we plot heat capacity as C/T versus $T^2$ at low temperatures as shown in the inset of Fig. 4. A fit of the type $C/T = \gamma + \beta T^2 + AT^2\exp(-\delta/T)$ (where the first term represents the electronic contribution, second the lattice contribution and third the anisotropic antiferromagnetic spin wave contribution with constants A and $\delta$] yields $\gamma$=0.7 mJ/mole-$K^2$ per formula unit and $\beta$=0.00094. This value of $\beta$ corresponds to $\Theta_D$ = 255 K. The electronic specific heat coefficient, $\gamma$, is a measure of the electronic density of states at the Fermi level. The near zero value of $\gamma$ in $La_2RuO_5$ may be contrasted with those of metallic and superconducting [2] $Sr_2RuO_4$ ($\gamma$=39 mJ/mol-$K^2$) and metallic ferromagnet [7] $SrRuO_3$ ($\gamma$=30 mJ/mol-$K^2$). The small value of $\gamma$ in $La_2RuO_5$ is consistent with the semiconducting nature of this compound (as inferred from electrical resistivity measurements presented below) where one does expect nearly zero or very low value for $\gamma$. Since the Ru-Ru magnetic interaction in $La_2RuO_5$ is expected to occur through the strong Ru-O hybridization, this may lead to some delocalization of the electrons and hence a small finite $\gamma$ value.

The electrical resistivity ($\rho$) of $La_2RuO_5$ was measured as a function of temperature (T) in the temperature range of 300-120K. The room temperature resistivity of this compound is about 700 $\Omega$ cm and increases with decreasing temperature (inset in Fig. 6). There is a gradual change in the slope of $\rho$-T curve near the magnetic transition which is seen more prominently in the ln$\rho$ versus 1/T plot shown in Fig. 6. The slope change in resistivity suggested the opening of a spin gap in this compound. The resistivity can be fitted to a thermally activated behaviour of the type $\rho=\rho_0\exp(E/kT)$ with slightly different values of activation energy E above and below the magnetic transition temperature. The activation energies obtained from the fit are 0.12eV above $T_N$ and 0.16eV below $T_N$.

These may be compared with the value of 0.28eV in semiconducting [8] $La_3RuO_7$ ($T_N$=20K). The large room temperature resistivity and the activation type of behaviour suggest that the compound $La_2RuO_5$ is semiconducting and that the carriers are highly localized.

When this work was completed, a literature search showed that a similar magnetic and resistivity behaviour has been observed in $La_4Ru_2O_{10}$ [9] which is identical to the presently investigated $La_2RuO_5$ (but for a factor of 2 in the formula) both having the same crystal structure though structural details are not presented in ref. 9. These authors have suggested that the magnetic ordering in $La_4Ru_2O_{10}$ is driven by ordering of the Ru orbitals as inferred from subtle changes in the crystal structure and Ru-O bond lengths from neutron diffraction studies. The ordering is accompanied by a large change in Ru moment also. The presence of magnetic ordering in $La_4Ru_2O_{10}$ or $La_2RuO_5$ is corroborated by our independent magnetic and resistivity measurements. However, the nature of the magnetic ordering remains unclear. No additional peaks have been observed [9] in the neutron diffraction pattern below the magnetic ordering temperature; while additional peaks would be expected for an antiferromagnetic ordering. It is possible that the Ru moments are considerably quenched in the magnetically ordered state and hence their presence is not seen through neutron diffraction measurements. Our heat capacity measurements show a large entropy change at the ordering temperature. The entropy change is found to be consistent with the ordering of $Ru^{4+}$ spin S=1 but may be fortuitous since contribution to entropy from structural transformation is not known. Our preliminary measurements [10] on $La_{2-x}Pr_xRuO_5$ system, for small values of x, reveal no structural change and a continued presence of the 170K transition. One would have expected the orbital ordering to be sensitive to the surroundings and likely to disappear with disorder but that does not appear to be the case. Clearly more work is necessary on this interesting compound.

Finally, it may be remarked that in most of the high temperature Cu-O based superconductors, the non-superconducting state is antiferromagnetic with a high $T_N$ arising from the Cu magnetic moments through the Cu-O hybridization. These compounds become superconducting on suitable dopings. It is possible that even $La_2RuO_5$, with very high magnetic ordering temperature, may exhibit superconductivity with appropriate doping.

# References

[‡] Present Address: Center for Superconductivity Research, University of Maryland, College Park, MD-20742

**Figure Captions**

Figure 1.  The Powder x-ray diffraction pattern of $La_2RuO_5$ sample refined by the Rietveld method using Fullprof software.

Figure 2.  Temperature variation of zero-field-cooled and field-cooled magnetization (M) for $La_2RuO_5$ in 50Oe applied magnetic field (H). Inset shows the M-H isotherms at various temperatures.

Figure 3.  Magnetic susceptibility ($\chi$) versus temperature (T) and $\chi^{-1}$ versus T for $La_2RuO_5$ in 5 kOe . The straight line in the $\chi^{-1}$–T data is the Curie-Weiss fit.

Figure 4.  Heat capacity (C) versus temperature (T) for $La_2RuO_5$. Inset shows the plot of C/T versus $T^2$ in the low temperature region and a fit to the equation $C/T=\gamma+\beta T^2+AT^2(-\delta/T)$.

Figure 5.  Magnetic contribution to heat capacity ($C_{mag}$) plotted as $C_{mag}/T$ versus T and the entropy S as a function of temperature for $La_2RuO_5$.

Figure 6.  Electrical resistivity ($\rho$) of $La_2RuO_5$ plotted as $\ln\rho$ versus 1/T and a fit to the activation type of behaviour. Inset shows the plot of electrical resistivity ($\rho$) versus temperature (T) for the same.

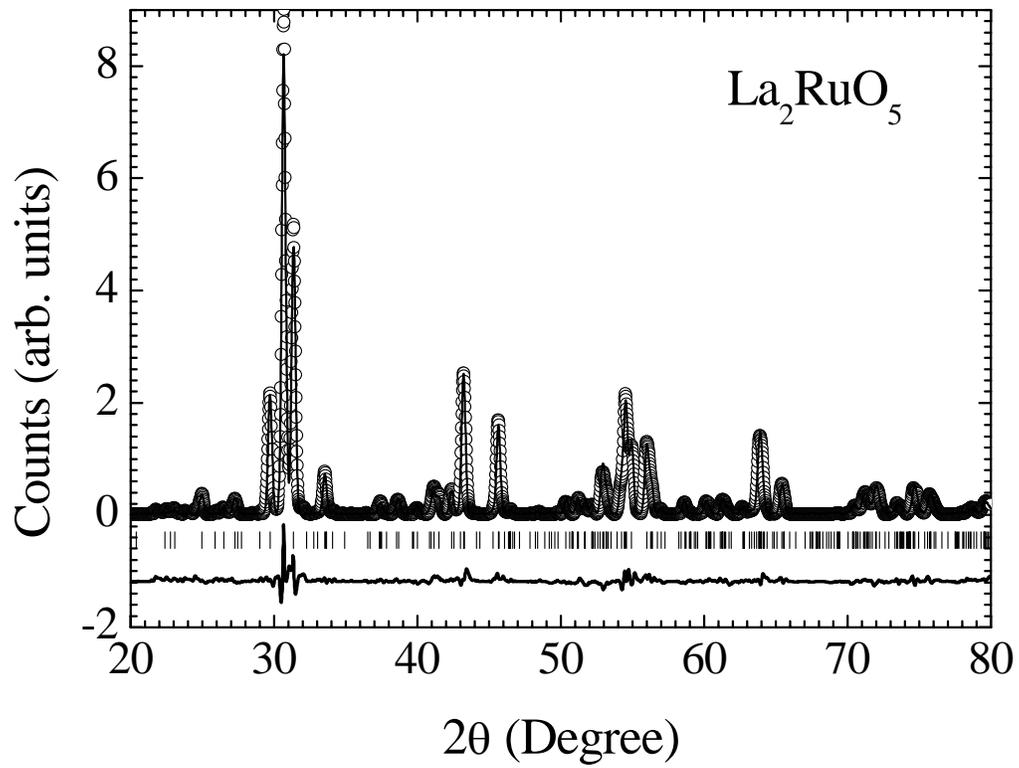

Figure 2 S.K. Malik *et al.*

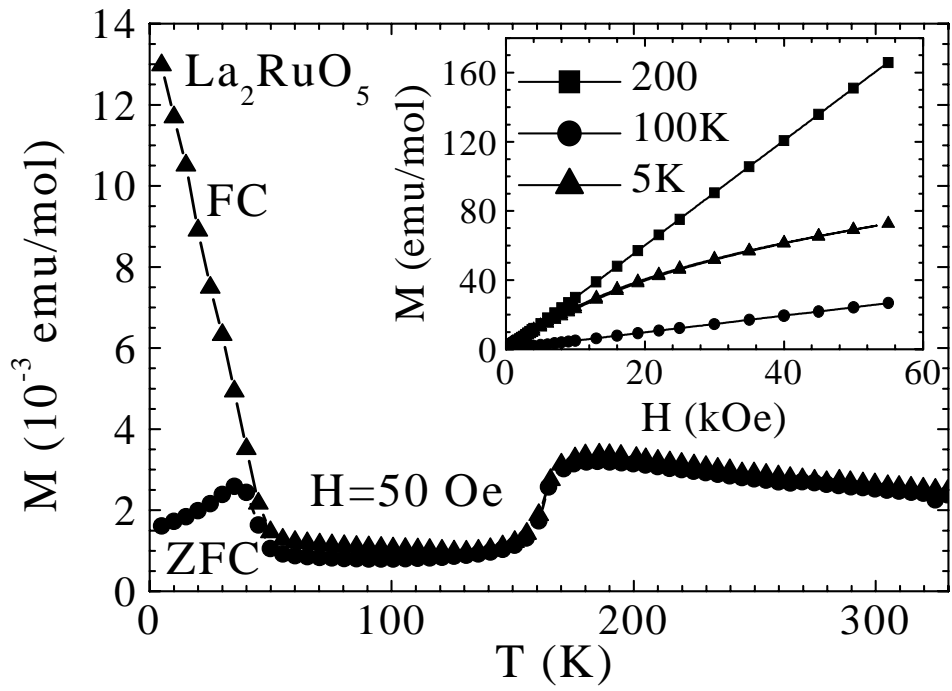

Figure 3 S.K. Malik *et al.*

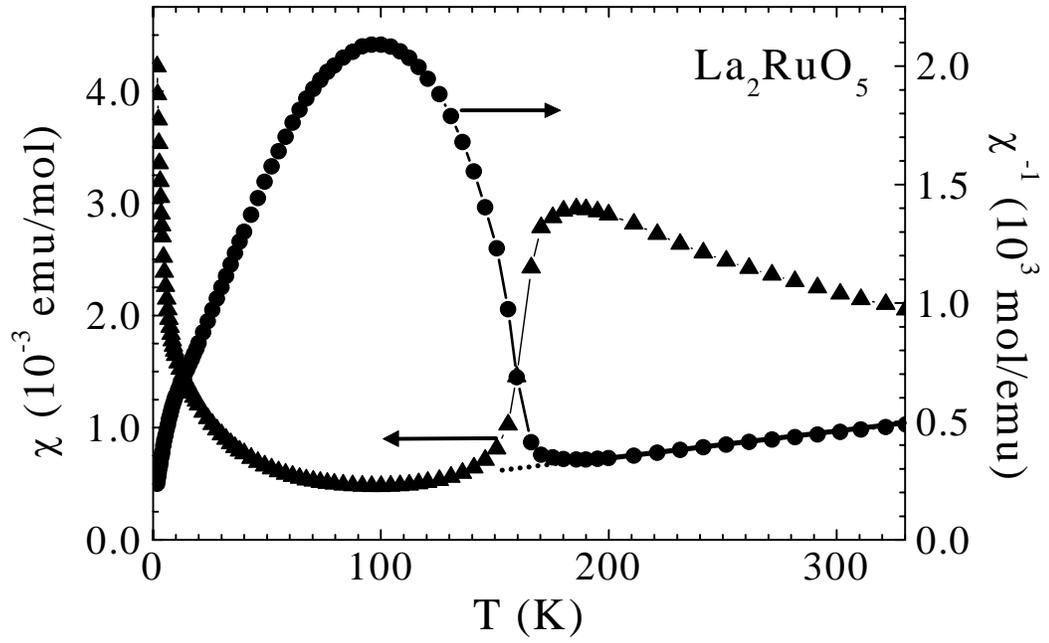

Figure 4 S.K. Malik *et al.*

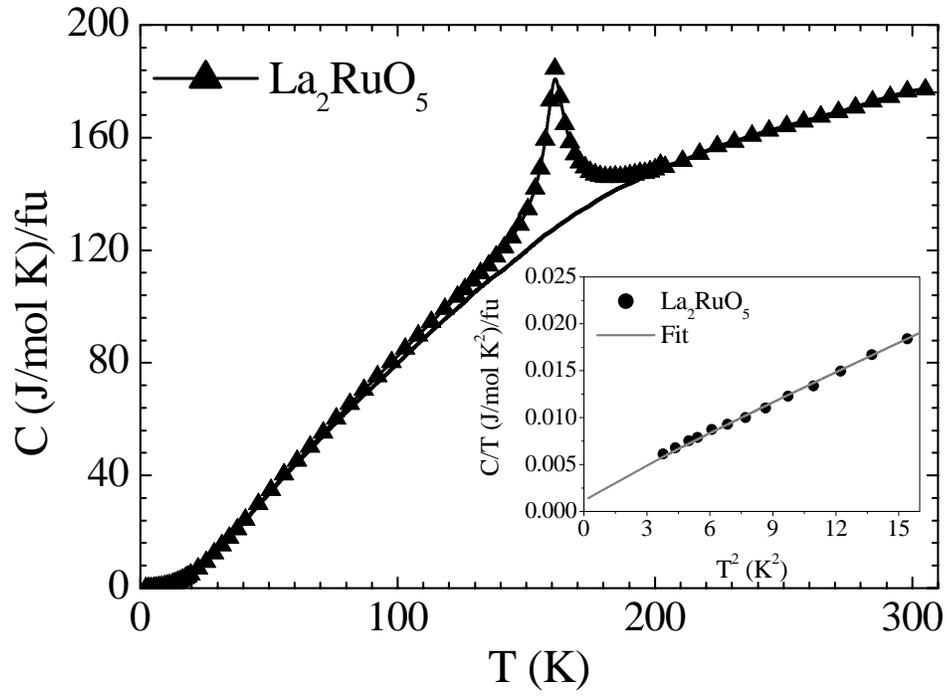

Figure 5 S.K. Malik *et al.*

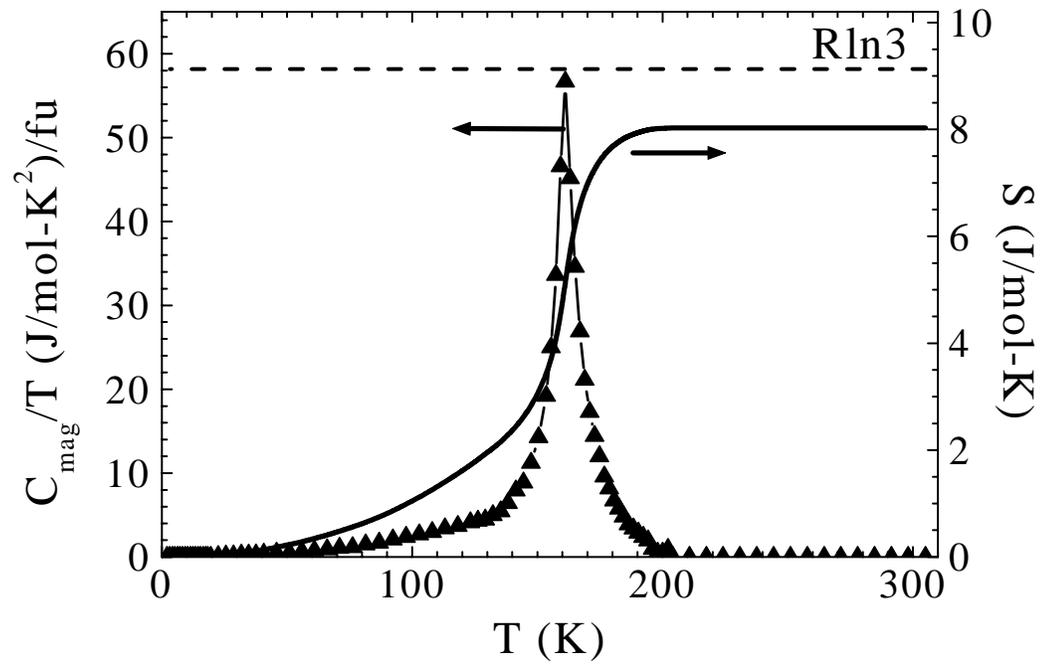

Figure 6 S.K. Malik *et al.*

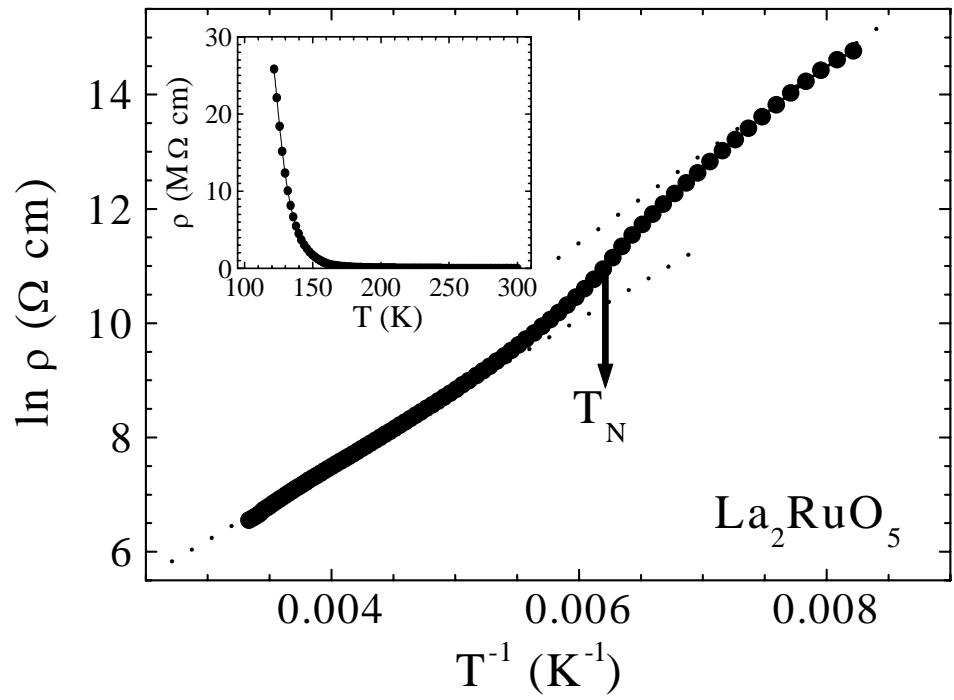